\documentclass[%
preprint,
amsmath,amssymb,
aps,
floatfix,
]{revtex4-2}
\usepackage{graphicx}
\usepackage{dcolumn}
\usepackage{bm}
\usepackage{hyperref}

\usepackage{multirow}
\usepackage{graphicx}
\usepackage{xcolor}
\usepackage{cancel}

\begin{document}

\title{Emergent broadband polarization entanglement from electronic and phononic Stokes--anti-Stokes indistinguishability}

\author{Diego Sier$^{1,4}$}
\author{Lucas Valente$^{1,4}$}
\author{Tiago A. Freitas$^{2}$}
\author{Marcelo F. Santos $^{3}$}
\author{Carlos H. Monken $^1$}
\author{Raul Corr\^ea$^{1,2}$}
\author{Ado Jorio$^{1}$}\email{adojorio@fisica.ufmg.br}

\affiliation{$^1$Departamento de F\'isica, Universidade Federal de Minas Gerais, Belo Horizonte, MG 30123-970, Brazil}
\affiliation{$^2$IDOR/Pioneer Science Initiative, Rio de Janeiro, RJ 22281-010, Brazil}
\affiliation{$^3$Instituto de F\'isica, UFRJ, Rio de Janeiro, RJ 21941-972, Brazil}
 \affiliation{$^4$These two authors contributed equally.}

\date{\today}

\begin{abstract}
Recently [{\it PRA} {\bf 108}, L051501 (2023)], it was shown that in a centrosymmetric cubic system, two photons from a broadband intense laser field can be converted into a pair of Stokes and anti-Stokes (SaS) entangled photons.
While the previous work was based on symmetry arguments, here we present a fully quantum theory for the SaS scattering that properly explains, quantitatively describes, and provides a means to predict its spectral and polarization properties (for diamond).
We also explore the possibilities offered by such system,
designing an entanglement map based on changes in the light-matter system. In particular, we show how the broadband polarization entanglement, that emerges from the interference between electronic and phononic degrees of freedom in the SaS scattering, depends on parameters such as Stokes-anti-Stokes Raman shift, scattering geometry and laser bandwidth, opening the avenue of exploration of such phenomenon in information processing.
\end{abstract}

\maketitle

\section{Introduction}

Photons produced in Raman scattering have their frequency shifted in relation to the incoming light ($\omega_L$), being called Stokes if their frequency is red-shifted ($\omega_S$) and anti-Stokes if they are blue-shifted ($\omega_{aS}$).
The two of them can be produced as part of a higher order process, which results in their correlation ($\hbar \omega_L + \hbar \omega_L' = \hbar \omega_S +\hbar \omega_{aS} $), with the coupled quanta being known as the Stokes--anti-Stokes (SaS) photon pair \cite{klyshko,parra,kishore2021}.
The operation of converting incoming photon(s) into different outgoing photon(s) in a medium is dictated by its susceptibility $\chi$ \cite{bloembergen,boyd},
and the $\chi^{(3)}$ enables two photons from a broadband intense laser field to interact with the medium to produce the correlated SaS photon pair.
In centrosymmetric materials, the second-order susceptibility is null and the non-linear response is usually dominated by the third-order tensor \cite{chi_3_tensor}.
While the non-classical correlations between SaS photons have been widely established \cite{lee2011entangling,lee2012macroscopic,Kasperczyk,saraiva,timsina2024resonant}, the presence of polarization entanglement in such process has been demonstrated only very recently, in diamond \cite{art_tiago}.

This recent result indicates that the SaS polarization entanglement comes from the indistinguishability between electronic (e-SaS) and phononic (p-SaS) mediated processes inside the centrosymmetric material.
It should, therefore, be possible to tailor the SaS state by playing with state preparation parameters, such as the scattering geometry, the SaS Raman shift and the laser bandwidth \cite{art_levenson,art_levenson2,filomeno_ass}.
In fact, the interplay of the SaS scattering properties should generate a map of different degrees of polarization entanglement depending on the balance between the electronic and phononic degrees of freedom.
Properly demonstrating, quantifying and exploring these possibilities is the goal of this paper. To achieve this, this work presents and experimentally certifies a fully quantum theory that describes the broadband scattered state in the SaS process, enabling a whole new level of the engineering of SaS photon pair generation.

Our article is organized as follows.
In Sec. II we provide the theory to support our experiments in which the two-photon quantum state arising in the Stokes--anti-Stokes scattering is derived.
Section III describes the experimental methods and their results, including the measurement of the Raman tensor components via spectral measurements for different photon polarization pairs and crystal angles, which is obtained with our theory.
In Sec. IV, the polarization entanglement of the SaS photon pairs is discussed, and we construct entanglement maps showing where to find maximum entanglement within the experimental parameters.
We conclude the article in Sec. V.

\section{Two-photon S\MakeLowercase{a}S quantum state}

The correlated SaS scattered intensity can be obtained by the third-order electric polarization in the material \cite{bloembergen, boyd}
\begin{equation}
    P_{i} (\omega_{aS})
    \propto
    \chi^{(3)}_{ijkl} (-\omega_{aS}, \omega_L, \omega_L', -\omega_S)
    \mathcal{E}_j (\omega_L) \mathcal{E}_k (\omega_L') E_l^\dagger (\omega_S),
\end{equation}
where $\mathcal{E}$ and $E$ represent the laser and scattered modes, respectively, and indexes $\{i,j,k,l\}$ represent polarization directions.
The laser modes are occupied by an intense classical field, which is unaffected by the scattering of a few photons, while the Stokes and anti-Stokes modes start in the vacuum.
The incident laser spectral amplitude is written as
$\mathcal{E}_j(\omega_L) = \mathcal{E}_{0j} G(\omega_L)$, with $G(\omega_L)$ describing the laser spectral distribution with $\int_0^\infty |G(\omega_L)|^2 d\omega_L = 1$.

The total SaS susceptibility is composed of an electronic (e-SaS) and a phononic (p-SaS) component~\cite{boyd, art_levenson2, bloembergen},
$\chi^{(3)}_{ijkl} = \chi^{(3)E}_{ijkl} + \chi^{(3)R}_{ijkl}$,
with $E$ for electronic and $R$ for Raman phononic.
The diamond structure belongs to the $O_h$ point group \cite{art_levenson,art_levenson2},
in which case $\chi^{(3)}_{ijkl}$ has only four independent terms~\cite{chi_3_tensor}, which are $\chi^{(3)}_{xxxx}, \chi^{(3)}_{xxyy}, \chi^{(3)}_{xyxy}, \chi^{(3)}_{xyyx}$ (assuming light propagating in the $z$ direction and that the $x$ and $y$ directions coincide with the crystallographic axes).
The frequency dispersion of the electronic contribution e-SaS can be neglected when it is very far from any electronic resonance, so that $ \chi^{(3) E}_{ijkl} (-\omega_{aS}, \omega_L, \omega_L', -\omega_S) = A^E_{ijkl} \delta(\omega_L +\omega_L' -\omega_S -\omega_{aS})$
becomes a complex constant tensor fulfilling energy conservation.
Ideally, the ratio between the e-SaS components in centrosymmetric materials is $\chi^{(3) E}_{xxxx} \approx 3 \chi^{(3) E}_{xyyx} \approx 3 \chi^{(3) E}_{xxyy} \approx 3 \chi^{(3) E}_{xyxy}$~\cite{bloembergen}.
However, experimental conditions are rarely ideal and we will only take these values as a reference.

The p-SaS susceptibility tensor, on the other hand, is known to be
$\chi^{(3) R}_{ijkl} \propto \sum_{\sigma} (\alpha^R_{ij,\sigma}\alpha^R_{kl,\sigma} + \alpha^R_{ik,\sigma}\alpha^R_{jl,\sigma} )$ \cite{art_levenson},
where $\alpha^R_{ij,\sigma}$ is the polarizability Raman tensor that describes the scattering of an incident electric field polarized at $i$ into a scattered mode polarized at $j$, via a phonon $\sigma$.
The Raman active vibrational mode of diamond belongs to the $T_{2g}$ irreducible representation of the $O_h$ point group.
Due to the relatively small probability of formation of SaS pairs, the spectral features of the phononic contribution to the susceptibility can be calculated from perturbation theory \cite{bloembergen} as being
\begin{equation}
    \chi^{(3) R}_{ijkl} (-\omega_{aS}, \omega_L, \omega_L', -\omega_S)
    =
    A^R_{ijkl}
    \frac{
        \gamma }
    { \omega_{ph} -\omega +i\gamma/2 }
    \delta(\omega_L +\omega_L' -\omega_S -\omega_{aS}).
\end{equation}
The tensorial part of the susceptibility is $A^R_{ijkl}$,
and he frequency dependence is composed of a Lorentzian-shaped amplitude of the Stokes scattering phonon frequency, $\omega = \omega_L -\omega_S$, around the resonance $\omega_{ph}$ with width $\gamma/2$ ($\gamma$ is the phonon decay rate, inversely proportional to its lifetime).
This susceptibility is the same as when the scattered fields are classical \cite{boyd,bloembergen}, being an extrapolation of the stimulated Raman scattering to the regime where the Stokes stimulation is turned off.

The probability amplitude that describes the SaS pair scattered at frequencies $\omega_S$ and $\omega_{aS}$ in a polarization pure state $|\psi\rangle$ is obtained by calculating
$
\langle 0|
        E_{l} (\omega_{S})
        E_{i} (\omega_{aS})
    | \psi \rangle
+ \langle 0|
        E_{i} (\omega_{aS})
        E_{l} (\omega_{S})
    | \psi \rangle
$
\cite{smith2006}, and since the electric polarization frequency component yields a scattered field in that same frequency and polarization,
the two-photon scattering amplitude probability at the SaS frequencies reads
\begin{eqnarray}\label{eq:SaS}
    \Psi_{li}(\omega_S, \omega_{aS})
    =
    C \mathcal{E}_{0j} \mathcal{E}_{0k}
    \int_{-\omega_S}^\infty
        G (\omega_S +\omega) G (\omega_{aS} -\omega) \nonumber\\*
    \times \left[
        A^E_{ijkl}
        + A^R_{ijkl}
            \frac{
                \gamma }
            { \omega_{ph} -\omega +i\gamma/2 }
    \right]
    d\omega .
\end{eqnarray}

The SaS scattering amplitude is widened by the laser bandwidth, which adds an energy uncertainty to the scattered photons in both electronic and phononic processes.
The factor $C$ in Eq.~(\ref{eq:SaS}) contains the efficiency of the SaS scattering in polarizations $l$ (Stokes) and $i$ (anti-Stokes), and depends on the scattered field frequencies and the scattering angles.
However, we can consider $C$ a constant in our experiments because we only collect pairs in forward scattering and because the laser bandwidth and phonon spectrum, taken separately in Eq.~(\ref{eq:SaS}), dominate any other frequency dependence.
Assuming a Gaussian amplitude spectrum for the laser, $G(\omega_L) = (\pi W^2)^{-1/4} e^{-(\omega_L -\omega_c)^2 / 2W^2}$, centered around the angular frequency $\omega_c$ with width $W$ (the laser power spectrum FWHM is $2\sqrt{\text{ln} 2} W/(2\pi)$),
the integral (\ref{eq:SaS}) can be solved analytically.
By separating the e-SaS and p-SaS parts, $f^E \equiv e^{-\bar{\omega}^2/W^2}$ and
\begin{equation}\label{eq:fR}
    f^R
    \equiv
    e^{-\bar{\omega}^2/W^2}
        \frac{\gamma e^{ -(\Omega -i\gamma/2)^2 / W^2 }}
        {2i \sqrt{\pi}W}
        \text{erfc} \left(
            \frac{\gamma}{2W} +i \frac{\Omega}{W}
        \right),
\end{equation}
with $\bar{\omega} \equiv (\omega_S +\omega_{aS})/2 -\omega_c$ and $\Omega \equiv (\omega_{aS} -\omega_{S})/2 -\omega_{ph}$,
then $\Psi_{li} = C \mathcal{E}_{0j} \mathcal{E}_{0k} (A^E_{ijkl} f^E + A^R_{ijkl} f^R)$.
Note that Eq.~(\ref{eq:fR}) comes from the integral in Eq.~(\ref{eq:SaS}), and the complex denominator in the Raman term yields the complex argument in the error function, associated with the phonon decay.
Also note that in $f^R$ the $e^{-\bar{\omega}^2/W^2}$ factor implies that detections at $\omega_S + \omega_{aS} = 2\omega_c$ are favored, while the dependence with $\Omega$ means that what dictates the spectrum is the SaS combined frequency $(\omega_{aS} - \omega_S)/2$ around $\omega_{ph}$.

With the probability amplitude (\ref{eq:SaS}) we can construct the SaS two-photon state generated by the scattering process.
We assume the laser polarization to be vertical (V) with respect to the laboratory, denoted by $\hat{\mathbf{e}}_V$.
If the crystallographic axis $x$ is also along the vertical direction, then  $\hat{\mathbf{e}}_V = \hat{\mathbf{x}}$, and the susceptibility $\chi^{(3)}_{ixxl}$ will govern the photon pair production.
In general, however, if the crystallographic axis $x$ and the laser polarization have an angle $\theta$ between them, such that $\hat{\mathbf{e}}_V = \cos(\theta) \hat{\mathbf{x}} + \sin(\theta) \hat{\mathbf{y}}$,
there will be a coherent sum of certain components of the susceptibility tensor defining the scattered photons polarization.
By writing the laser field components as $\mathcal{E}_{0x} = \mathcal{E}_0 \cos\theta$
and $\mathcal{E}_{0y} = \mathcal{E}_0 \sin\theta$,
the scattered (non-normalized) state can be written as
\begin{eqnarray}
    |\Psi_{SaS}\rangle (\omega_S, \omega_{aS}; \theta)
    =
    [ Y^E_{VV}(\theta) f^E
        + Y^R_{VV}(\theta) f^R ] |VV\rangle \nonumber\\*
    + [ Y^E_{HH}(\theta) f^E
        + Y^R_{HH}(\theta) f^R ] |HH\rangle \nonumber\\*
    + [ Y^E_{VH}(\theta) f^E
        +Y^R_{VH}(\theta) f^R ] (|VH\rangle + |HV\rangle). \label{eq:SaS-theta}
\end{eqnarray}
The functions $Y^\eta_{li} (\theta)$ depend on the angle $\theta$ and the $A^\eta_{ijkl}$ tensor components,
\begin{subequations}\label{eq:Y-factors}
    \begin{eqnarray}
        Y^\eta_{VV} (\theta) &\equiv& C\mathcal{E}_{0}^2 [
            (\mathcal{S}^4 +\mathcal{C}^4) A^\eta_{xxxx} \nonumber\\*
             & &+ 2 \mathcal{S}^2\mathcal{C}^2 (A^\eta_{xyyx} + A^\eta_{xxyy} + A^\eta_{xyxy})
        ],
    \end{eqnarray}
    \begin{eqnarray}\label{eq:YHH}
        Y^\eta_{HH} (\theta) &\equiv& C\mathcal{E}_{0}^2 [
            2\mathcal{S}^2\mathcal{C}^2 A^\eta_{xxxx}
            + (\mathcal{S}^4 +\mathcal{C}^4) A^\eta_{xyyx} \nonumber\\*
            && -2\mathcal{S}^2\mathcal{C}^2 (A^\eta_{xxyy} + A^\eta_{xyxy})
        ],
    \end{eqnarray}
    \begin{eqnarray}
        Y^\eta_{VH} (\theta) &\equiv& C \mathcal{E}_{0}^2
        (\mathcal{S}^2 -\mathcal{C}^2) \mathcal{S}\mathcal{C} \nonumber\\*
            && \times \left(
            A^\eta_{xxxx} - A^\eta_{xyyx} - A^\eta_{xxyy} - A^\eta_{xyxy}
        \right),
    \end{eqnarray}
\end{subequations}
where $\mathcal{S} \equiv \sin\theta$ and $\mathcal{C} \equiv \cos\theta$.
Equation (\ref{eq:SaS-theta}) means that the angle $\theta$ drives the balance between the contribution of the e-SaS spectrum in $f^E$ and the p-SaS spectrum in $f^R$,
and this balance will be different for each pair of scattered photons polarization $VV$, $HH$ and $VH$ (equal to $HV$), depending on how the tensor components are combined.
If a different, possibly more complicated, laser spectrum were used, a new spectral function $f^R$ would result, but the state of Eq.~(\ref{eq:SaS-theta}) would still have the same form.

The state in Eq.~(\ref{eq:SaS-theta}) is, in general, entangled in polarization.
In particular, for $\theta = 0^\circ$, $Y^\eta_{VH} (0^\circ) = 0$, and $\{|VV\rangle, |HH\rangle\}$ form a Schmidt basis \cite{nielsen_chuang},
meaning that if the amplitudes of these vector components are equal, state $|\Psi_{SaS}\rangle$ (\ref{eq:SaS-theta}) is maximally entangled, while if one of them is zero, it is separable.
In this sense, the problem of producing a state with a high amount of polarization entanglement becomes a matter of tailoring the relative amplitude of the two components of this Schmidt basis.

To obtain a prediction for the SaS state generated as a function of $\theta$, one needs to characterize the tensor components $A^\eta_{ijkl}$ in the functions $Y^\eta_{li} (\theta)$, Eq.~(\ref{eq:Y-factors}).
The functions $f^\eta$ in Eq.~(\ref{eq:SaS-theta}) are known and only depend on the laser central frequency $\omega_c$, bandwidth $W$, the phonon frequency $\omega_{ph}$ and decay rate $\gamma$, which are given experimental conditions.
Measuring the relative scattered intensities into $|HH\rangle$ and $|VV\rangle$ in $\theta = 0^\circ$ and $\theta= 45^\circ$ (angles in which there are no $|VH\rangle$ and $|HV\rangle$ components) allows us to retrieve the values of $A^\eta_{ijkl}$.
Next, we present an experiment that was used to extract these quantities in diamond, and show how they translate into SaS entanglement.

\section{Experimental characterization of the centrosymmetric S\MakeLowercase{a}S scattering}

\subsection{Setup and methods}

\begin{figure}[ht]
    \centering
    \includegraphics[scale=0.30]{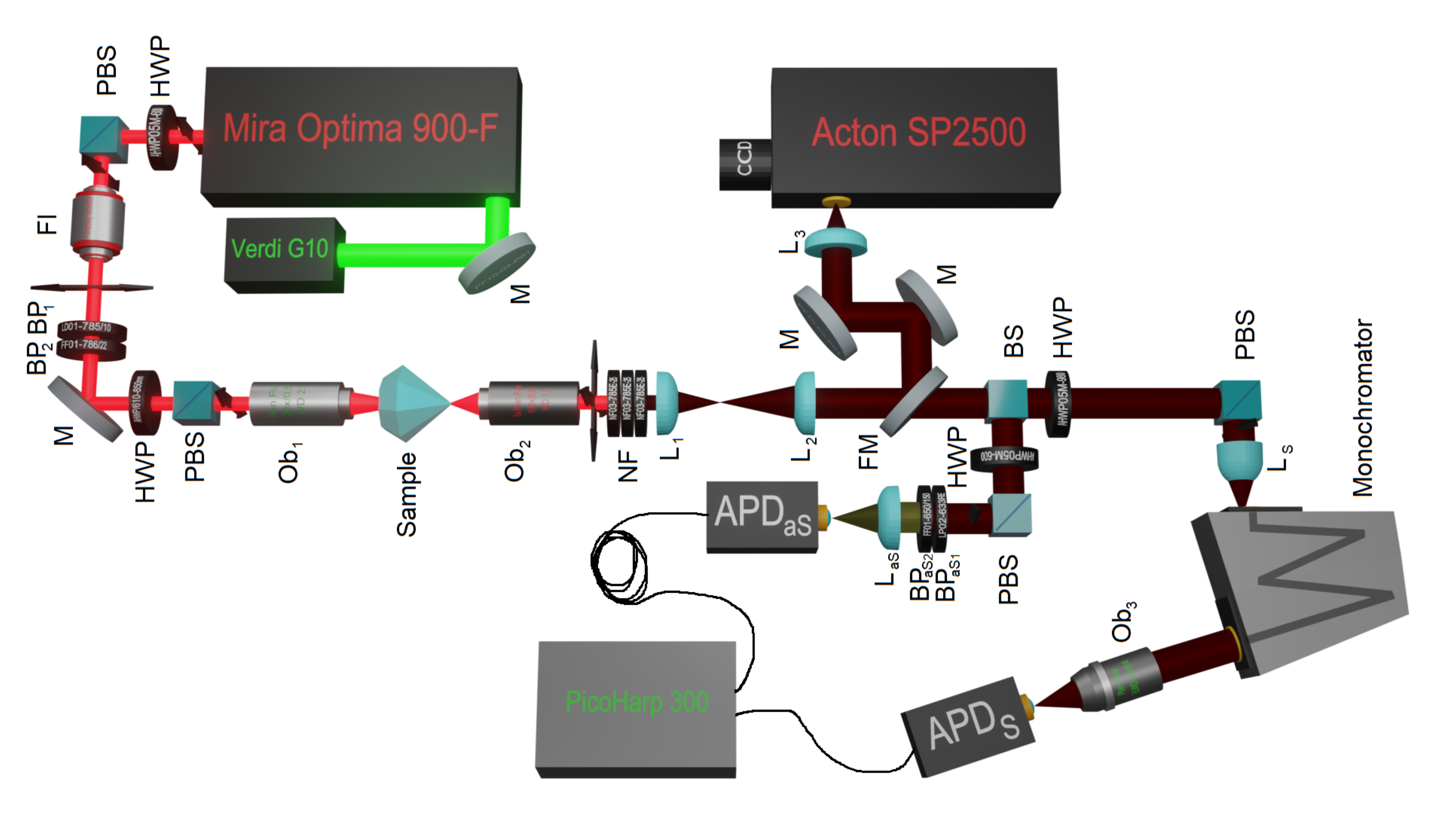}
    \caption{Depiction of experimental configuration. (M) mirror, (HWP) half-wave plate, (PBS) polarization beam splitter, (FI) faraday isolator, (BP$_1$, BP$_2$) interference bandpass filter, (Ob$_1$) objective for laser focusing on the sample (20x, N.A. 0.50, W.D. 2.1mm), (Ob$_2$) objective for collecting scattering signals (100x, N.A. 0.90, W.D. 1.0mm), (NF) notch filter, (L$_1$, L$_2$, L$_3$, L$_{aS}$) convergent plano-convex achromatic lenses, (FM) flip mirror, (BS) 50/50 beamsplitter, (BP$_{aS_1}$, BP$_{aS_2}$) interference bandpass filter to anti-Stokes, (L$_S$) aspheric condenser convergent lens, (Ob$_3$) objective for laser focusing on APD$_S$ (20x, N.A. 0.45, W.D. 8.2-6.9 mm).}
    \label{IM:setup}
\end{figure}

Figure \ref{IM:setup} schematically describes the experimental setup used. A Verdi G-10 laser at 532 nm wavelength excites a Ti:Sapphire crystal from a Mira 900F laser, which generates a train of pulses at 76 MHz repetition rate, with interval between pulses of 13.6 ns and 130 fs nominal pulse width. The output signal, tuned at 781 nm wavelength with $104$ mW of power, passes through a half-wave plate (HWP) and a polarization beam splitter (PBS) to control the power and polarization before being focused on the diamond sample.
The transmitted Raman signal is collected and a flippable mirror can direct this signal for two distinct optical paths.
With the mirror flipped on the path, the signal is sent to a spectrometer and the Raman spectrum of the sample is obtained with a Charge Coupled Device (CCD).
Once the mirror is flipped out of the optical path, the sample signal is divided in a 50/50 beamsplitter and each branch is separately sent through a HWP and a PBS, which are used to select the polarization of the photons.
After polarization selection, in one of the branches a bandpass filter (FF01-650/150 - Semrock) and a longpass filter (LP02-633RE - Semrock) are used to filter the anti-Stokes photons that are then focused and detected in an avalanche photodiode (APD).
In the other branch, the Stokes signal is filtered with a monochromator and then focused with an objective lens in an APD.
The electric signals from the APDs (SPCM-AQRH-14 - Excelitas) are connected in a time-correlated single photon counting apparatus (PicoHarp 300 - PicoQuant) and a histogram of the temporal difference between the signals is created.

The sample used was a diamond grown by a CVD process (Type IIac, 100-oriented, from Almax) positioned so that the laser propagates in the (001) direction of the crystal.
The sample assembly allows it to be rotated on the laser propagation axis such that the angle between the laser polarization and the crystallographic axes of the sample ($\theta$) can be varied.
The polarization of the laser is vertical (V) with respect to the laboratory and the angle $\theta$ is set as $\theta = 0^{\circ}$ for the (100) crystallographic axis of the sample matched with the vertical direction.
For each monochromator position (each selected $\delta\omega = \omega_c - \omega_S$), one histogram of temporal difference is obtained in $300$ seconds of acquisition. This scanning procedure is done for two orientations of the sample ($\theta = 0^{\circ}$ and $45^{\circ}$), and for each sample orientation the spectrum is obtained with the polarization of the photons incident on the avalanche photodiode (APD) detectors being selected in two ways: a spectrum with Stokes and anti-Stokes photons with vertical polarization (VV) and a spectrum with both photons with horizontal polarization (HH).
Details of the treatment of the experimental data can be found in Appendixes~\ref{app:experimental_data} and \ref{app:efficiency}.

\subsection{Results}

\begin{figure}[ht]
    \centering
    \includegraphics[scale=0.085]{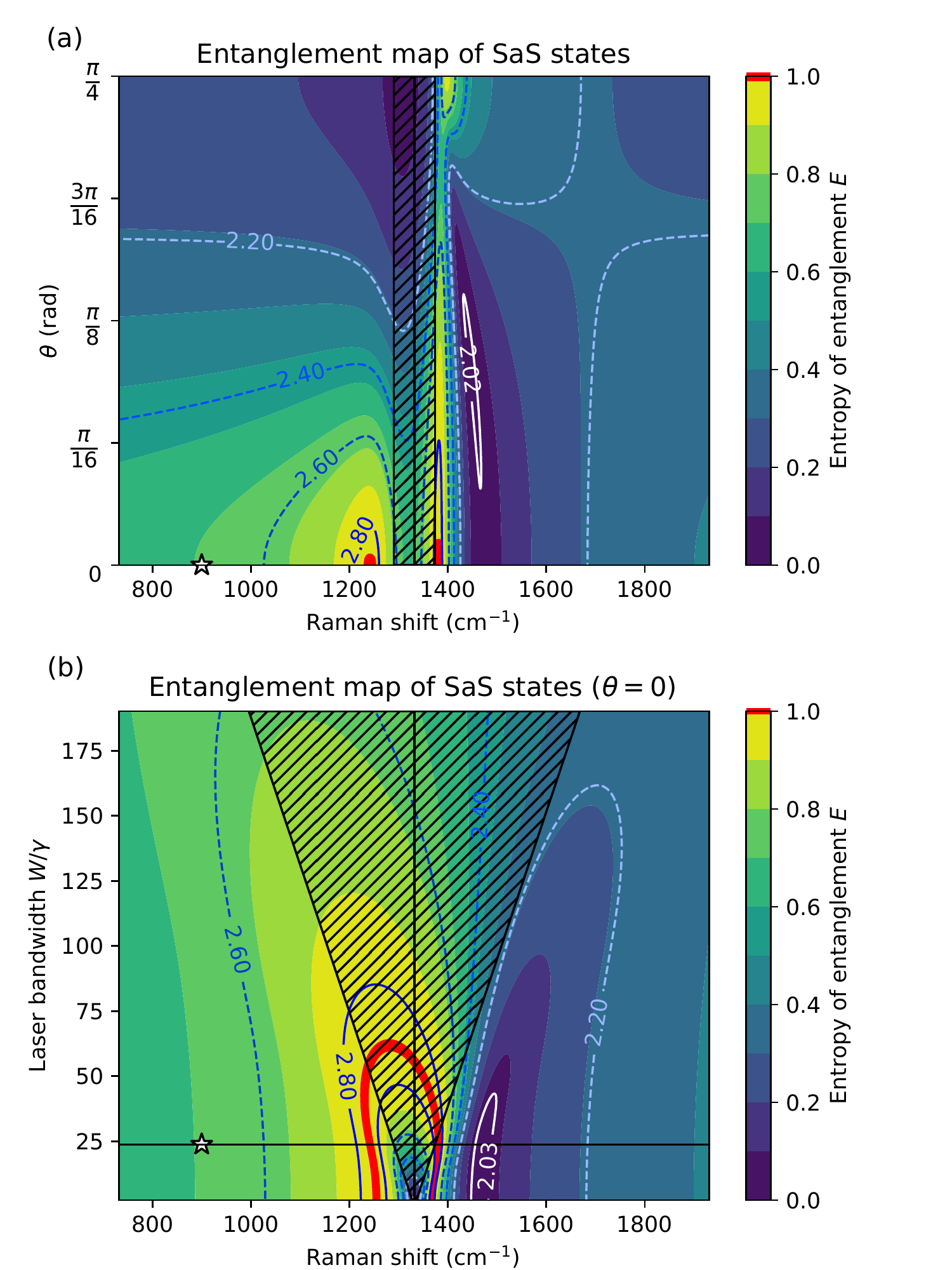}
    \caption{(a,b) Time-correlated SaS photon pairs intensity
    at VV (blue dots) and HH (red dots) polarization,
    for crystallographic orientation at (a) $\theta=0^\circ$
    and (b) $\theta=45^\circ$.
    The $I_{SaS}^{corr}$ data was corrected to account for system efficiency and adjusted by the square of the laser power (see Appendix~\ref{app:efficiency}),
    and the pump laser central wavelength used was 781 nm.
    Lines are the theoretical model: the flat curves are the mean counts $\langle I_{SaS}^{corr}\rangle$ at VV$(0^\circ)$ and HH($45^\circ$);
    for the resonance curves, HH$(0^\circ)$ and VV($45^\circ$), $\omega_{ph}/2\pi = 1332 \text{ cm}^{-1}$,
    $\gamma = 11 \text{ cm}^{-1}$,
    $\omega_c/2\pi = 12.7 \times 10^3 \text{ cm}^{-1}$,
    and $W/2\pi = 42 \text{ cm}^{-1}$
    in Eq.~(\ref{eq:SaS}).
    The fitting $Y$ factors from Eq.~(\ref{eq:SaS-theta}) were
    $Y^E_{HH}(0^\circ) = \langle I_{SaS}^{corr} (\text{VV}(0^\circ)) \rangle^{1/2} (0.68-i0.12)$,
    $Y^E_{VV}(45^\circ) = \langle I_{SaS}^{corr} (\text{VV}(0^\circ)) \rangle^{1/2} (1.61-i0.55)$,
    and $Y^R_{HH}(0^\circ) = Y^R_{VV}(45^\circ) = 51450\ [\text{Hz/W}^2]^{1/2}$. (c,d) g$^{(2)}(0)$ values for a diamond crystallographic orientation of (c) 0° and (d) 45°.
    The large error bars in $g^{(2)}(0)$ reflect the errors in the low accidental counts (see Appendix~\ref{app:experimental_data}).
    The light gray lines indicate the classical limit g$^{(2)}(0)=2$.}
    \label{IM:Fig}
\end{figure}

The intensity spectra of correlated SaS pairs for VV (blue dots) and HH (red dots) polarization at $\theta=0^\circ$ and $\theta=45^\circ$ are displayed in Figs.~\ref{IM:Fig}~(a) and (b), respectively.
This experiment was specifically designed by combining the single-wavelength polarization analysis presented in \cite{art_tiago} with the unpolarized spectral analysis provided in \cite{filomeno_ass}, to prove the strength of our theoretical approach.
According to group theory analysis, VV(0$^\circ$) configuration does not exhibit a Raman contribution ($A^R_{xxxx} = 0$ in Eq.~(\ref{eq:Y-factors})) and only involves non-resonant electronic transitions. For this reason, the blue curve shown in Fig.~\ref{IM:Fig}~(a) corresponding to this configuration is flat.
Additionally, Fig.~\ref{IM:Fig}~(a) shows a dark blue solid line, obtained by averaging the intensity values $I_{SaS}^{corr}$ for this experimental configuration.
We multiply this value by the monochromator slit spectral width $\Delta\omega/2\pi = 11 \text{ cm}^{-1}$, yielding $\langle I^{corr}_{SaS} (VV(0^\circ)\rangle = (27.5 \pm 3.5)\times 10^3 \text{ Hz/W}^2$.
Since $Y^E_{VV}(0^\circ) = C\mathcal{E}_0^2A^E_{xxxx}$ and in our model it is proportional to $\langle I^{corr}_{SaS} (VV(0^\circ))\rangle^{1/2}$, we have a reference value for $A^E_{xxxx}$.

To compare the measured spectra with our spectral functions $f^E$ and $f^R$, Eq.~(\ref{eq:fR}), one must take into account the width of the frequency filters.
The experimental setup (Fig.~\ref{IM:setup}) contains a bandpass filter in the anti-Stokes branch, and a monochromator in the Stokes one.
The anti-Stokes filter is modelled as a square frequency filter from anti-Stokes Raman shift of 900 cm$^{-1}$ to 3000 cm$^{-1}$,
and the Stokes frequencies contributing to the counts are the ones that pass the monochromator slit, also modelled as a square frequency filter, but with width $\Delta\omega/2\pi = 11 \text{ cm}^{-1}$ around the Stokes central frequency $\omega_S$.
We thus evaluate
\begin{equation}\label{eq:filters}
    I^{\text{det}}_{li} (\omega_S)
    =
   \int_{\omega_S-\Delta\omega/2}^{\omega_S+\Delta\omega/2}
    \left[ \int_{F_\text{aS}} |\Psi_{li}(\omega_S',\omega_{aS}')|^2 d\omega_{aS}' \right]
    d\omega_S' ,
\end{equation}
where $F_{\text{aS}}$ represents the aS filter frequencies,
and $\Psi_{li}(\omega_S,\omega_{aS})$ is the wave function in Eq.~(\ref{eq:SaS}).

By looking at the other relevant scattering geometries, we obtain the other tensorial components of Eq.~(\ref{eq:Y-factors}) in relation to $A^E_{xxxx}$.
The HH(45$^\circ$) configuration also exhibits only e-SaS, as seen in the red data in Fig~\ref{IM:Fig}~(b). The mean and standard deviation of the counts gives $(1.60 \pm 0.28) \times 10^3$ Hz/W$^2$.
The HH(0$^\circ$) data in Fig.~\ref{IM:Fig}~(a) (red points) contain p-SaS with the characteristic resonance signature, which comes from $A^R_{xyyx} \neq 0$ in Eq.~(\ref{eq:Y-factors}b), contributing with a Raman scattering in the $|HH\rangle$ polarization.
Using $Y^E_{HH}(0^\circ) = \langle I_{SaS}^{corr} (\text{VV}(0^\circ)) \rangle^{1/2} (0.68-i0.12)$
and $Y^R_{HH}(0^\circ)=51450\ [\text{Hz/W}^2]^{1/2}$ fits the data with the shown theoretical curve in Fig.~\ref{IM:Fig}~(a).
The VV(45$^\circ$) data is the last configuration left, and it is shown in Fig.~\ref{IM:Fig}~(b) in blue.
To fit it,
we fix the Raman factor to be the same as in the HH(0$^\circ$) configuration, $Y^R_{VV}(45^\circ) = 51450\ [\text{Hz/W}^2]^{1/2}$,
and use $Y^E_{VV}(45^\circ) = \langle I_{SaS}^{corr} (\text{VV}(0^\circ)) \rangle^{1/2} (1.61-i0.55)$.
With the above values and working with Eq.~(\ref{eq:Y-factors}) we obtain the electronic and Raman $A^\eta_{ijkl}$ components, summarized in Table \ref{tab:suscep}.
Details of the calculations can be found in Appendix \ref{app:tensor_components}.
\begin{table}[h]
    \caption{List of measured electronic and Raman tensor components ratios $A^\eta_{ijkl}/A^E_{xxxx}$.}
    \centering
    \begin{tabular}{c|c}
    \hline \hline
        Tensor component    &   Value \\
    \hline
       $A^E_{xyyx}/A^E_{xxxx}$  & $(0.37 \pm 0.04) +i(-0.07 \pm 0.01)$ \\
       $(A^E_{xxyy} + A^E_{xyxy})/A^E_{xxxx}$  & $(0.89 \pm 0.10) +i(-0.07 \pm 0.01)$ \\
       $A^R_{xxxx}/A^E_{xxxx}$  & $0 $ \\
       $A^R_{xyyx}/A^E_{xxxx}$  & $(171 \pm 12) $ \\
       $(A^R_{xxyy} + A^R_{xyxy})/A^E_{xxxx}$  & $(171 \pm 12) $ \\
    \hline \hline
    \end{tabular}
    \label{tab:suscep}
\end{table}

For completeness the second-order correlation function $g^{(2)}(0)$ is evaluated
with the ratio $[I^{corr}_{SaS}(\Delta \tau =0) +\bar{I}_{SaS}(\Delta \tau \neq 0)]/\bar{I}_{SaS}(\Delta \tau \neq 0)$,
where $I^{corr}_{SaS}(\Delta\tau = 0)$ are the measurements in Fig.~\ref{IM:Fig}~(a) and (b), and $\bar{I}_{SaS}(\Delta \tau \neq 0)$ accounts for the uncorrelated pair production. A plot of $g^{(2)}(0)$ is shown in Fig.~\ref{IM:Fig}(c,d),
where it is evident that near the resonance the values are the lowest, accounting for the high number of uncorrelated SaS photon pairs produced.
The large error bars in $g^{(2)}(0)$ reflect the errors in the low accidental counts (see Appendix~\ref{app:experimental_data}), and the theory agrees well with the data in spite of the error size.
Furthermore, the curves for HH(0$^\circ$) and VV(45$^\circ$) show a drop in $g^{(2)}(0)$, being higher below and lower above the resonance region, while the VV($0^\circ$) and HH($45^\circ$) keep roughly the same value.
The asymmetry in HH(0$^\circ$) and VV(45$^\circ$) is associated with the uncorrelated counts being symmetrical with respect to the resonance peak, and thus the asymmetry in the correlated counts for these configurations,
which contain a p-SaS contribution, is transferred to the $g^{(2)}(0)$ curves. This asymmetry is a result of the constructive ($\delta \omega < \omega_{ph}$) versus destructive ($\delta \omega > \omega_{ph}$) interference between the e-SaS and the p-SaS, which explains the Cooper-pair-like behavior of $I_{SaS}^{corr}$ \cite{saraiva,filomeno_ass}.
On the other hand, the correlated counts in the VV(0$^\circ$) and HH(45$^\circ$) configurations are a purely e-SaS contribution, and thus are a flat curve,
leading to an overall symmetric $g^{(2)}(0)$.

\section{Polarization entanglement in the S\MakeLowercase{a}S pairs}

With the values of $A^\eta_{ijkl}$ in hand, one can predict what is the polarization entanglement in the SaS scattered state $|\Psi_{SaS}\rangle (\theta)$ of Eq.~(\ref{eq:SaS-theta}) for given values of $\omega_S$ and $\omega_{aS}$.
We use, as an entanglement measure, the entropy of entanglement $E = -\text{Tr}_i(\rho_i \text{log}_2 \rho_i)$ of either subsystem $i = \{S, aS\}$ with reduced state $\rho_i$ \cite{nielsen_chuang}, which, since our global state is pure, is zero for separable states and unity for maximally entangled states.

\begin{figure}[ht]
    \centering
    \includegraphics[scale=0.55]{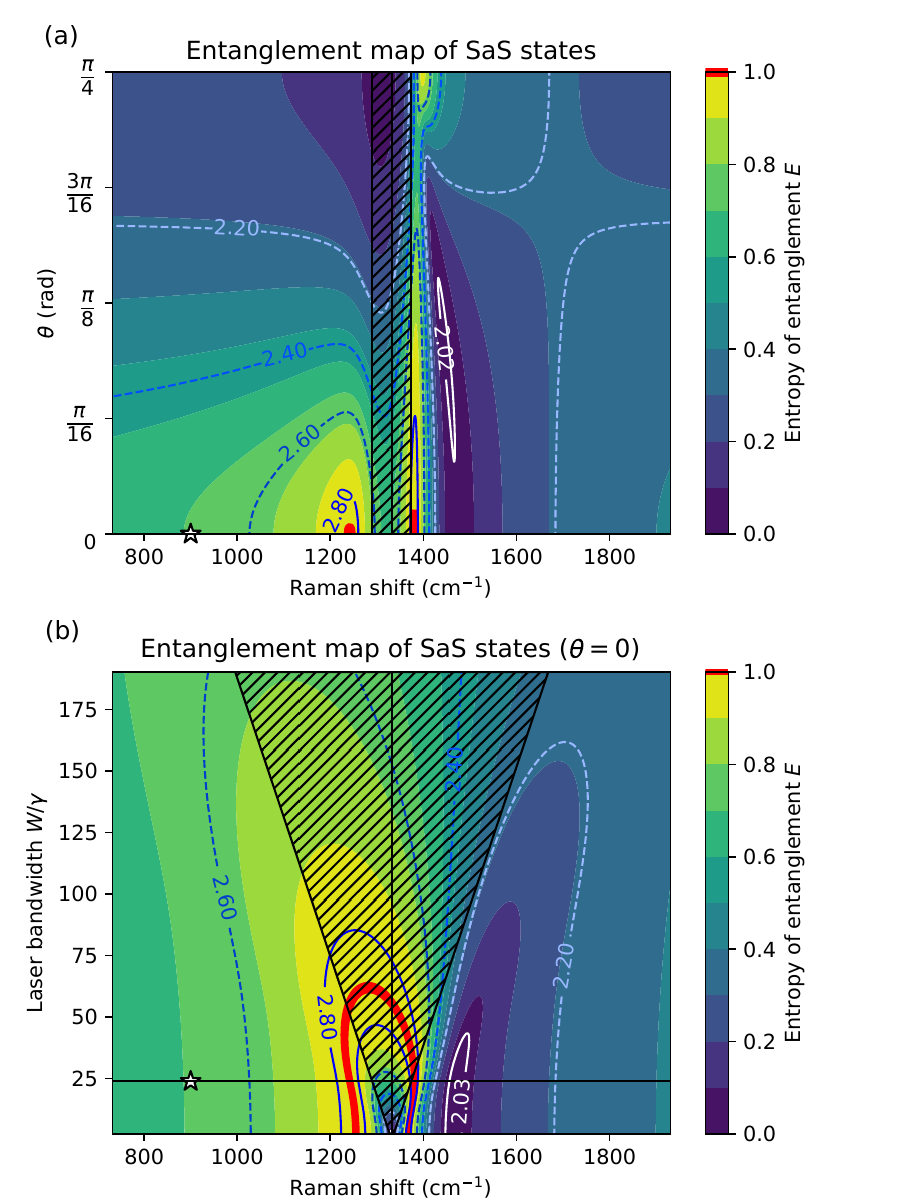}
    \caption{Entanglement map for the SaS states in (a) the present experiment and (b) a generic centrosimetric cubic system.
    In (a), the laser bandwidth $W$ is fixed at $W = 264 \text{ cm}^{-1}$, the value of our experiment, as localized in (b) by a horizontal line.
    The black vertical line is the resonant Raman shift $\omega_{ph}/2\pi$, and the red regions mark the entanglement maxima ($E > 0.999$).
    As tensor components in Eq.~(\ref{eq:Y-factors}) we use the values from Table \ref{tab:suscep}.
    The star localizes the experiment of Ref.~\cite{art_tiago}.
    In blue shades, we plot contour curves for the Gisin parameter $F$.
    The hatched regions, within $2W$, indicate a high production of independent SaS photon pairs.
    }
    \label{IM:entanglement}
\end{figure}

In Fig.~\ref{IM:entanglement} we show a contour plot of $E(\omega_S,\omega_{aS};\theta, W)$,
under the condition that $\omega_{aS} = 2\omega_c - \omega_S$, that is, a symmetric SaS Raman shift.
On the horizontal axis we vary the Raman shift, going through the resonance in diamond at $\omega_{ph}/2\pi = 1332$\,cm$^{-1}$, identified by a black vertical line,
and on the vertical axis we plot (a) the angle $\theta$ between the laser linear polarization and the crystallographic axis,
going from $\theta=0^\circ$ to $45^\circ$ (due to the symmetry of the crystal, all other $\theta$ values will be related to this range),
and (b) the laser bandwidth $W$, going from zero to $W = 190\gamma$, and where the value for our experiment $W = 24 \gamma$ ($W = 264 \text{ cm}^{-1}$ or FWHM 70 cm$^{-1}$) is indicated by a black horizontal line.
The value of $\gamma=11 \text{ cm}^{-1}$ is obtained experimentally using a CW laser.
The thick red regions indicate the parameters where $E \rightarrow 1$, showing maximum entanglement.

Maximum entanglement occurs at $\theta = 0^\circ$,
in a region below resonance close to 1200 cm$^{-1}$ and another above, close to 1400 cm$^{-1}$.
The measurements in Fig.~\ref{IM:Fig} illustrate this for the $0^\circ$ and $45^\circ$ conditions, for which $|HH\rangle$ and $|VV\rangle$ are a Schmidt basis, and the balance between $HH$ and $VV$ counts reflects how close it is to maximum entanglement.
At $0^\circ$ the purely electronic response (VV) is strong, and maximum entanglement occurs where the HH response crosses it with the same amount of scattered photon pairs.
Conversely, at 45$^\circ$ the purely electronic response (HH) is weak in comparison with the VV scattering, which contains the Raman response,
thus making it impossible for the probability amplitude of the $|HH\rangle$ component to balance with the $|VV\rangle$ one, when it would reach a maximally entangled state.
For other crystal angles $\theta$, comprising the middle section of Fig.~\ref{IM:entanglement}~(a), the Schmidt basis can be determined for any $\theta$ using our model of Eq.~(\ref{eq:SaS-theta}).
Measurements on this basis allows one to experimentally obtain the SaS pair entanglement entropy for any desired crystal angle without the need for a complete state tomography.

If the laser bandwidth $W$ grows, as shown in Fig.~\ref{IM:entanglement}~(b), the shape of the resonance curve is smoothed out, so the ratio between the $|VV\rangle$ and $|HH\rangle$ coefficients gets more and more even along the SaS spectrum.
Because of this, the entanglement extremes become less pronounced as $W$ grows, until the point that it is not possible to reach maximum entanglement anymore, at $W \approx 70\gamma$.
On the other hand, it also gets harder to obtain a separable state, which depends on a peak of either the $|VV\rangle$ or the $|HH\rangle$ component in relation to the other.

In Fig.~\ref{IM:entanglement} there is a hatched region indicating where the uncorrelated SaS pair production is high (low $g^{(2)}$ region in Fig.~\ref{IM:Fig}~(c,d)), corresponding to $2W$, around $1.2 \times$ FWHM.
In this region, the scattered state is not properly represented by Eq.~(\ref{eq:SaS}), but it needs to be complemented by other SaS scattering events, which involves the scattering of real phonons, that happens within a laser bandwidth around the Raman resonance \cite{filomeno_life}.
Outside this region, though, state (\ref{eq:SaS}) is a good representation of the scattered SaS state, and our entanglement measure is representative.

To complete the analysis, we also draw level curves for the violation of Bell-type CHSH inequalities as a function of the light-matter parameters. These curves correspond to the so-called Gisin parameter $F$ \cite{gisin91, chefles97} which, for pure states, reads
$F = 2(1 + 2\mathcal{P})^{1/2}$, where $\mathcal{P} = 1 - \text{Tr}_i(\rho_i)^2$ is the linear entropy of either subsystem $i$. For separable states $F=2$, which means that no separable state can go above the classical upper bound in the CHSH inequality, equal to 2.
It reaches its maximum at $F= 2\sqrt{2} \approx 2.83$ for maximally entangled states.
We plot the $F$ parameter in Fig.~\ref{IM:entanglement} as blue shaded contour curves, and its smallest values within a white contour. Note that the maximal violation coincides with the maximally entangled states near 1200 cm$^{-1}$ and 1400 cm$^{-1}$. Since Bell analysis is state dependent, and the generated $|\Psi_{SaS}\rangle$ state depends on the values of the light-matter parameters, the appropriate Bell angles to reach the maximum violation change for different regions of the map. As an example, the violation in the maximally entangled states near 1200 cm$^{-1}$ is obtained for linear polarization angles $(0, \pi/4)$ for one of the photons and $(\pi/8, 3\pi/8)$ for the other, where the angles are in relation to $V$ polarization.
Note that the quality of the entanglement, characterized by both the Gisin parameter $F$ and the entanglement entropy $E$, are independent of the rate of pair generation, so no matter how many SaS photon pairs are generated, our analysis yields how much entanglement will be available with each pair.

Finally, we can localize the result of Ref. \cite{art_tiago} in Fig.~\ref{IM:entanglement}, represented by a white star.
In the paper, a state close to a Bell state at a symmetric Raman shift of 900 cm$^{-1}$ with $\theta = 0^\circ$ was found, with amplitude ratio of $|HH\rangle$ to $|VV\rangle$ at $\sqrt{0.28/0.72} \approx 0.62$.
Calculating the ratio with our state (\ref{eq:SaS-theta}) yields 0.49, and the discrepancy
comes from our anti-Stokes filter cutting some of the Raman contribution in HH at 900 cm$^{-1}$ (left edge of Fig.~\ref{IM:Fig}~(a)), which does not happen in Ref.~\cite{art_tiago}.
This region can be seen in the bottom left of Fig.~\ref{IM:entanglement}~(a) and (b), where $E \approx 0.7$ and $F \approx 2.5$.

\section{Conclusion}

This work presents a fully quantum theory that describes the broadband scattered state in the SaS process, including photon pair polarization.
It incorporates experimental conditions like the laser bandwidth and crystal orientation, to allow for the calculation of any sought physical quantity (e.g. $g^{(2)}(0)$ and entanglement).
The efficiency of the process can be controlled by how close the system is to the resonance with a Raman active phonon, and it is a matter of state engineering to choose the appropriate laser--crystal angle and SaS frequency to obtain a good balance between a good amount of entanglement and a sufficient pair production rate.
This theory has been elusive to find despite a decade-long effort \cite{parra, saraiva, filomeno_ass, guimaraes2020, diaz2020, art_tiago}.
The result is explored here for diamond, but it should be general for other centrosymmetric media, including silicon, whose electronic band gap (1.12 eV) is one order of magnitude higher than the phonon energy ($< 0.1$ eV), respecting the assumptions of our model. This should considerably bolster the engineering of SaS photon pair generation.
We note that related results were independently obtained in Ref. \cite{vento2025}.

Other good candidates for SaS photon pair generation which can feed on the principles of our theory are two-dimensional materials.
Graphene, however, is limited by a strong luminescence that prevents photon correlation studies \cite{jorio2014, parra}, and in transition metal dichalcogenides (TMDs) the phonon energies are too low, which demands low cryogenic temperatures to avoid a high rate of thermally excited anti-Stokes photons.

For nonzero $\chi^{(2)}$ materials, spontaneous parametric down-conversion (SPDC) enables broadband pair generation as well, as has been recently demonstrated \cite{javid2021}.
However, in SPDC the generation of a polarization entangled photon pair is due to the decay of a single highly energetic photon \cite{boyd}, demanding a UV laser for the generation of photon pairs in the visible spectrum, while the use of four-wave mixing (FWM) allows the generation of pairs with frequencies close to any excitation laser.
This peculiarity means that using FWM can have more versatility in the choice of optics for entangled photon pair generation.

\section*{Acknowledgments}
We acknowledge Professor C. Galland and his team for insightful discussions.
This work was supported by IDOR/Pioneer Science Initiative (www.pioneerscience.org),
Conselho Nacional de Desenvolvimento Cient\'ifico (INCT-IQ 465469/2014-0, 302872/2019-1, 421469/2023-4, 307619/2023-0),
Funda\c c\~ao de Amparo \`a Pesquisa do Estado de Minas Gerais (APQ-01860-22, APQ-04852-23, RED0008123)
and Fundação Carlos Chagas Filho de Amparo à Pesquisa do Estado do Rio de Janeiro (CNE E-26/200.307/2023, E-26/210.069/2020).


%

\appendix

\section{Analysis of the experimental data}\label{app:experimental_data}

In this Appendix we provide the details of the experimental data and its analysis.
The experimental data is composed of a collection of temporal difference histograms associated with frequency filtered Stokes detections.
The Stokes signal, filtered with the monochromator, is scanned around the diamond's first order Raman peak (1332 cm$^{-1}$) in a spectral region ranged from 895 to 1750 cm$^{-1}$.
For each monochromator position, one histogram of temporal difference is obtained in $300$ seconds of acquisition time.
This scanning procedure is done for two positions of the sample ($\theta = 0^{\circ}$ and $45^{\circ}$) and for each sample position the spectrum is obtained with the polarization of the photons incident on the APDs being selected in two ways: a spectrum with Stokes and anti-Stokes photons with vertical polarization (VV) and a spectrum with both photons with horizontal polarization (HH).

From the histograms two kinds of experimental data are obtained, the intensity of correlated pairs $I^{corr}_{SaS}(\Delta \tau =0)$, and the second-order normalized correlation function $g^{(2)}(\tau=0)$, respectively given by
\begin{eqnarray}\label{eq:I_sas_corr_exp}
    I^{corr}_{SaS}(\Delta \tau =0)& =& I_{SaS}(\Delta \tau = 0) - \bar{I}_{SaS}(\Delta \tau \neq 0),\\
    \label{eq:g2_exp}
    g^{(2)}(\Delta\tau=0) &=& \frac{I_{SaS}(\Delta \tau = 0) }{\bar{I}_{SaS}(\Delta \tau \neq 0)},
\end{eqnarray}
where $ I_{SaS}(\Delta \tau = 0)$ is the total coincident counts of SaS photons pairs in $\Delta\tau = 0$ and $\bar{I}_{SaS}(\Delta \tau \neq 0)$ is the average of uncorrelated S,aS photons pairs in $\Delta\tau \neq 0$.
These data are obtained from each experimental point in the spectral range and corrected for the detection efficiency of the system (see Appendix \ref{app:efficiency}).

\begin{figure}[h!]
    \centering
    \includegraphics[scale=0.1]{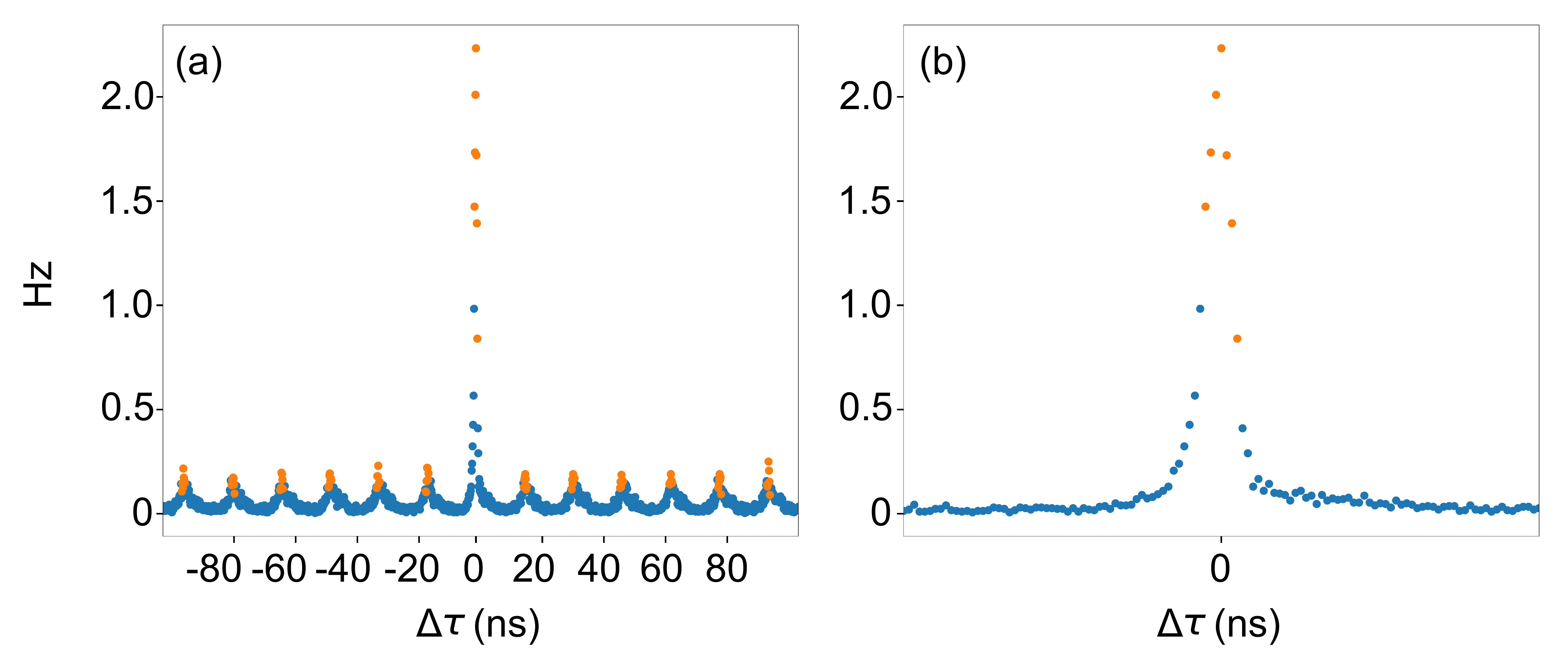}
    \caption{(a) Typical histogram for counts of SaS photon pairs created by the PicoHarp 300 is shown, with the utilized values for integrating the peaks highlighted in yellow. (b)  Peak for $\Delta \tau =0$ zoomed in to clarifying the yellow points.}
    \label{IM:histogram}
\end{figure}

The count of correlated Stokes--anti-Stokes (SaS) photon pairs was experimentally obtained using the setup shown in Fig.~\ref{IM:setup}. The PicoHarp 300 correlator provides a histogram of the detection rate of photon pairs as a function of time. An example is shown in Fig.~\ref{IM:histogram}. The peaks of coincidences are separated by 13 nanoseconds due to the time rate of the pulsed laser. With the 0.128 picosecond resolution of this correlator, and considering the time jitter of both APDs (Avalanche Photodiodes) at 350 picoseconds, we have three time bins of points before and after the peak value. This is type of the histogram integration we use to improve the statistics of the photon counts. All data values were obtained from a 300 seconds measurement.

For each data histogram, we determine the correlated rate of SaS photon pairs using the peak at $\Delta \tau = 0$, and the uncorrelated rate using the average of the peaks at $\Delta \tau \neq 0$.
For intensities of correlated SaS photon pairs, $I_{SaS}(\Delta \tau = 0)$, we can create a spectral graph of the obtained values, as shown in Fig.~\ref{IM:central_acd}~(a) and (b) for crystallographic orientations of 0° and 45°, respectively, with VV (blue dots) and HH (red dots) polarization.
The experimental values related to the average of the uncorrelated SaS photon pairs, $\bar{I}_{SaS}(\Delta \tau \neq 0)$, can be demonstrated in Fig.~\ref{IM:central_acd}~(c) for the sample oriented at 0° with VV and HH polarization (following the same color scheme as the other figures) and in Fig.~\ref{IM:central_acd}~(d) for the diamond oriented at 45°.
We observed that in all data the values of $I_{SaS}(\Delta \tau = 0)$ show a peak in the resonance region, which can be attributed to an increase in uncorrelated SaS photon pairs in this region, as shown in Fig.~\ref{IM:central_acd}~(c,d).

\begin{figure}[h!]
    \centering
    \includegraphics[scale=0.08]{Fig5.pdf}
    \caption{Experimental graphs without efficiency corrections for the intensity of correlation pairs, $I_{SaS}(\Delta \tau = 0)$, per second for (a) 0$^\circ$ and (c) 45$^\circ$ crystallographic orientations using VV and HH polarizations. The other graphs show the uncorrelated SaS photon pairs, $\bar{I}_{SaS}(\Delta \tau \neq 0)$, per second for VV polarization (blue dots) and HH polarization (red dots), for the crystallographic orientation of (c) 0$^\circ$ and (d) 45$^\circ$.  }
    \label{IM:central_acd}
\end{figure}

\begin{figure}[h!]
    \centering
    \includegraphics[scale=0.1]{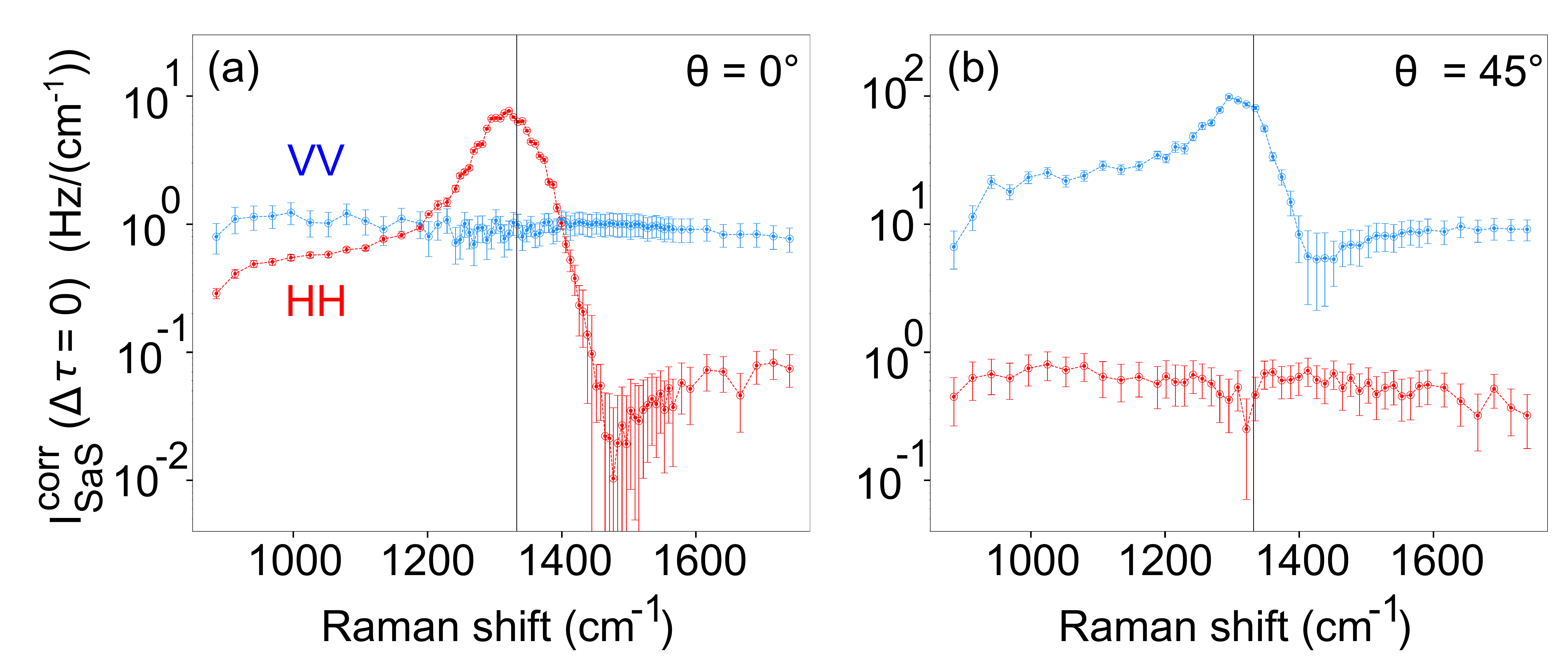}
    \caption{Graphs for the intensities of correlated SaS photon pairs, $I^{corr}_{SaS}(\Delta \tau =0)$, without efficiency corrections using VV and HH polarizations for (a) 0$^\circ$ and (b) 45$^\circ$ angular positions of diamond. }
    \label{IM:sem_cor}
\end{figure}

By subtracting $\bar{I}_{SaS}(\Delta \tau \neq 0)$ from $I_{SaS}(\Delta \tau = 0)$, we obtain the intensity of correlated SaS photons, $I^{corr}_{SaS}(\Delta \tau = 0)$. In Fig.~\ref{IM:sem_cor}, we show the curves for 0° (Fig.~\ref{IM:sem_cor}~(a)) and 45° (Fig.~\ref{IM:sem_cor}~(b)) for VV and HH polarizations. The values here differ from those shown in Fig.~\ref{IM:Fig} due to efficiency corrections made to the data, but the behavior of the curves remains the same.

The error bars of the uncorrelated counts (Fig.~\ref{IM:central_acd}~(c) and \ref{IM:central_acd}~(d)) were estimated in two steps,
by noting that the curves consist of a central Gaussian peak plus a constant background.
The error of the constant background is two times the standard deviation of the counts around the mean.
On the other hand, the error of the Gaussian is obtained by fitting a Gaussian curve to the data with the Python SciPy library, and then the error bars correspond to the uncertainty in the fit, provided by the library.

The error bars of the correlated counts in Fig.~\ref{IM:central_acd}~(a), red curve, and Fig.~\ref{IM:central_acd}~(b), blue curve, were obtained from the curves without the Raman contribution, namely VV at 0$^\circ$ and HH at 45$^\circ$, just like we did with the uncorrelated counts,
that is, separating the flat background and the Gaussian peak contribution.
Then we assume that the error of the HH or VV counts are of the same order at 0$^\circ$ and at 45$^\circ$,
that is, we use the same error bars of VV at 0$^\circ$ in the curve of Fig.~\ref{IM:central_acd}~(a) in the VV(45$^\circ$) data in Fig.~\ref{IM:central_acd}~(b).
Similarly, we use the same error bars of HH at 45$^\circ$ in the curve of Fig.~\ref{IM:central_acd}~(b) in the HH(0$^\circ$) data in Fig.~\ref{IM:central_acd}~(a).

The error bars for the correlated counts $I_{SaS}^{corr} (\Delta \tau =0) = I_{SaS} (\Delta \tau =0) - \bar{I}_{SaS} (\Delta \tau \neq 0)$
in Fig.~\ref{IM:Fig} were obtained by error propagation of independent variables from both
$\Delta \tau = 0$ counts $I_{SaS}(\Delta \tau = 0)$ of Fig.~\ref{IM:central_acd}~(a,b), and $\Delta \tau \neq 0$ counts $\bar{I}_{SaS}(\Delta \tau \neq 0)$ counts of Fig.~\ref{IM:central_acd}~(c,d),
according to
\begin{equation}
\delta I_{SaS}^{corr}(\Delta \tau =0) =
\sqrt{
    [\delta I_{SaS}(\Delta \tau =0)]^2
    + [\delta \bar{I}_{SaS}(\Delta \tau \neq 0)]^2
    }
    ,
\end{equation}
where $\delta x$ indicates the uncertainty of quantity $x$.
In Fig.~\ref{IM:sem_cor}, the errors are the same as the ones in Fig.~\ref{IM:Fig}~(a,b), but without efficiency correction.

The error bars of $g^{(2)}(0)$ in Fig.~\ref{IM:Fig}~(c,d) were obtained by standard error propagation of independent variables from the $\Delta \tau = 0$ counts $I_{SaS}(\Delta \tau = 0)$ of Fig.~\ref{IM:central_acd}~(a,b), and $\Delta \tau \neq 0$ counts $\bar{I}_{SaS}(\Delta \tau \neq 0)$ counts of Fig.~\ref{IM:central_acd}~(c,d),
with $g^{(2)}(0) = I_{SaS}(\Delta \tau =0)/\bar{I}_{SaS}(\Delta \tau \neq 0)$,
yielding
\begin{equation}
\delta g^{(2)}(0) =
    \sqrt{
    \left[ \frac{1}{\bar{I}_{SaS}(\Delta \tau \neq 0)} \right]^2
    [\delta I_{SaS}(\Delta \tau =0)]^2
    +
    \left[ \frac{I_{SaS}(\Delta \tau = 0)}{(\bar{I}_{SaS}(\Delta \tau \neq 0))^2} \right]^2
    [\delta \bar{I}_{SaS}(\Delta \tau \neq 0)]^2
    }
    .
\end{equation}

\section{Efficiency Correction}\label{app:efficiency}

\begin{figure}[h!]
    \centering
    \includegraphics[scale=0.095]{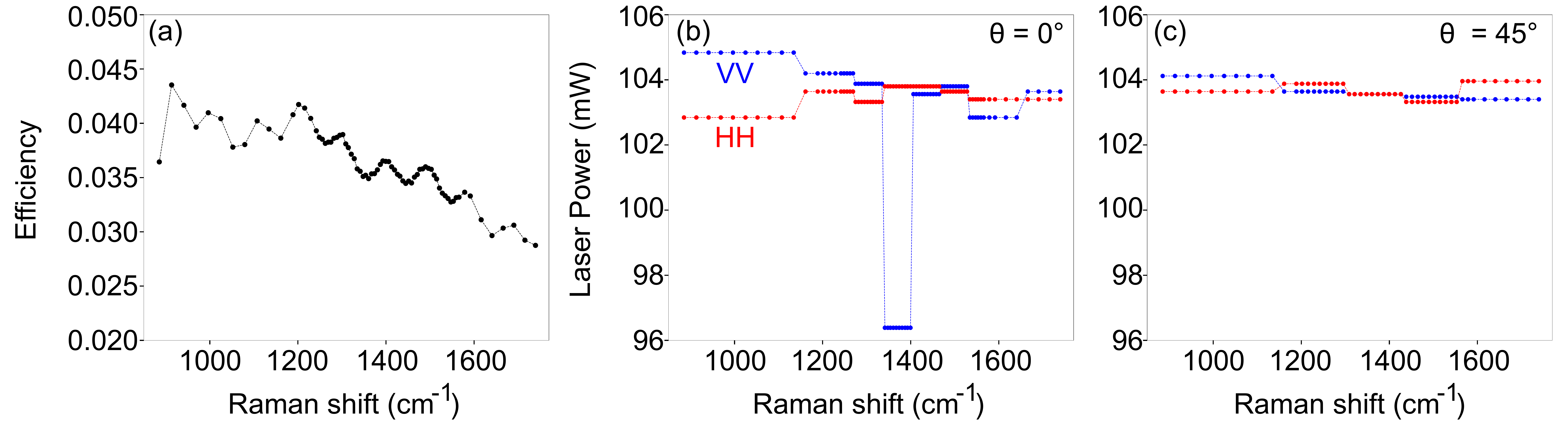}
    \caption{(a) Efficiency of the total optical elements used as a function of Raman Shift in the experimental setup of Fig.~\ref{IM:setup} (all values were obtained from the manufacturer's website). Moreover, this figure shows the spectrum of the laser power in miliwatt for (b) $\theta  =0^\circ$ and (c)  $\theta  =45^\circ$ for VV and HH polarizations.  }
    \label{IM:efficiency}
\end{figure}

Optical measurements are affected by the efficiency of the optical elements used in the experimental setup. Therefore, to ensure that the measured values are closer to an ideal measurement—i.e., without photon losses along the optical path and during detection—a correction was applied to the values in Fig.~\ref{IM:sem_cor}. This correction was based on the product of all the efficiencies of the optical elements used in the experiment, with the final value as a function of the Raman shift exemplified in Fig.~\ref{IM:efficiency}~(a). All efficiency values were obtained from the manufacturer's website.

In addition to this correction, the fact that fluctuations in laser power can occur was taken into account. Therefore, the laser power was measured every ten measurements. These values are specified in Fig.~\ref{IM:efficiency}~(b) for the data related to the crystallographic orientation of 0° for VV and HH, and in Fig.~\ref{IM:efficiency}~(c) for the 45° orientation for both polarizations.

From these values, the correction
\begin{equation}\label{eq:correction}
    \mbox{Real Value} = \frac{\mbox{Measured Value}}{\mbox{Efficency} \times \mbox{Laser Power}^2}
\end{equation}
is obtained for each measured value.
Note that in Eq.~\eqref{eq:correction} the laser power is squared due to the quadratic nature of SaS photon pair formation with the laser power.

\section{Obtaining the material tensor components}\label{app:tensor_components}

The values of the material tensor components in Table~\ref{tab:suscep} are calculated using the plot fitting parameters
$Y^E_{HH}(0^\circ) = \langle I_{SaS}^{corr} (\text{VV}(0^\circ)) \rangle^{1/2} (0.68-i0.12)$,
$Y^E_{VV}(45^\circ) = \langle I_{SaS}^{corr} (\text{VV}(0^\circ)) \rangle^{1/2} (1.61-i0.55)$,
and $Y^R_{HH}(0^\circ) = Y^R_{VV}(45^\circ) = 51450\ [\text{Hz/W}^2]^{1/2}$,
and Eq.~(\ref{eq:Y-factors}).

According to group theory analysis, the VV(0$^\circ$) configuration does not exhibit a Raman contribution ($A^R_{xxxx} = 0$ in Eq.~(\ref{eq:Y-factors})) and it only involves non-resonant electronic transitions.
For this reason, the blue curve shown in Fig.~\ref{IM:Fig}~(a) of the article corresponding to this configuration is flat.
Additionally, Fig.~\ref{IM:Fig}~(a) shows a dark blue solid line, obtained by averaging the intensity values $I_{SaS}^{corr}$ for this experimental configuration,
and the standard deviation of the experimental points around this mean is at 13\%.
With the mean and standard deviation of VV(0$^\circ$) counts, times the monochromator slit spectral width $\Delta\omega/2\pi = 11 \text{ cm}^{-1}$, at
$\langle I_{SaS}^{corr} (\text{VV}(0^\circ)) \rangle = (27.5 \pm 3.5)\times 10^3 \text{ Hz/W}^2$, we take its square root and use $\langle I_{SaS}^{corr} (\text{VV}(0^\circ)) \rangle^{1/2} = MY^E_{VV}(0^\circ) = (166 \pm 11)\ [\text{Hz/W}^2]^{1/2}$ as our reference value,
where $M = 0.551$ is a constant accounting for the integration of $f^E(\omega_S,\omega_{aS})$ over the filter frequencies (see Eq.~(\ref{eq:SaS-theta}) and  (\ref{eq:filters})).
The uncertainty is obtained by error propagation.
Since $Y^E_{VV}(0^\circ) = C\mathcal{E}_0^2A^E_{xxxx}$, we have a reference value for $A^E_{xxxx}$.

We can now look at the HH(0$^\circ$) data in Fig.~\ref{IM:Fig}~(a), represented by the red points, which contains p-SaS, as can be seen from the characteristic resonance signature.
In this configuration we have $A^R_{xyyx} \neq 0$ in Eq.~(\ref{eq:Y-factors}), contributing with a Raman scattering in the $|VV\rangle$ polarization, and which we will calculate.
To fit the data with Eqs. (\ref{eq:SaS})--(\ref{eq:Y-factors}), we used $Y^E_{HH}(0^\circ) = M Y^E_{VV}(0^\circ) (0.68 -i0.12)$
and $Y^R_{HH}(0^\circ) = 51450\ [\text{Hz/W}^2]^{1/2}$.
Since $Y^\eta_{HH}(0^\circ) = C\mathcal{E}_0^2A^\eta_{xyyx}$,
these numbers lead to
$$A^E_{xyyx}/A^E_{xxxx} = (0.37 \pm 0.04) +i (-0.07 \pm 0.01)$$
and
$$A^R_{xyyx}/A^E_{xxxx} = (171 \pm 12).$$
Note that $|A^E_{xyyx}|/|A^E_{xxxx}|$ is close to the $1/3$ expected from group theory [13] and we have a 3\% absorption in the electronic degrees of freedom.
Also, the electronic-to-Raman ratio $A^E_{xyyx}/A^R_{xyyx} \approx 0.002$ should be comparable to the susceptibilities ratio near the resonance since, by $\chi^{(3) R}_{ijkl} = A^R_{ijkl} \gamma / (\omega_{ph} - \omega_L+\omega_S +i\gamma/2)$, at resonance $|\chi^{(3) R}_{ijkl}| = 2 |A^R_{ijkl}|$.

The HH(45$^\circ$) configuration also exhibits only e-SaS, as seen in the red data in Fig~\ref{IM:Fig}~(b) of the main article.
The mean and standard deviation of the counts gives $(1.60 \pm 0.28) \times 10^3 \text{ Hz/W}^2$, which means a 17\% fluctuation.
Taking the square root of the counts give us
$MY^E_{HH}(45^\circ) = (40 \pm 4)\ [\text{Hz/W}^2]^{1/2}$.
Using $MY^E_{HH}(45^\circ) = MC\mathcal{E}_0^2[A^E_{xxxx} +A^E_{xyyx} -(A^E_{xxyy} +A^E_{xyxy}) ]/2$
and the values obtained with the VV(0$^\circ$) and HH(0$^\circ$) data, we find
$$(A^E_{xxyy} +A^E_{xyxy})/A^E_{xxxx} = (0.89 \pm 0.10) +i (-0.07 \pm 0.01).$$
This value does not coincide with the $2/3$ expected, suggesting that there is some unaccounted effect, which Ref. [13] attributes to some two-photon absorption processes.
Furthermore, for the p-SaS contribution in this configuration to be zero, that is, $Y^R_{HH}(45^\circ) = 0$, we need $(A^R_{xxyy} +A^R_{xyxy}) = A^R_{xyyx}$ (see Eq.~(\ref{eq:YHH})).

The VV(45$^\circ$) data is the last configuration left, and it is shown in Fig.~\ref{IM:Fig}~(b) of the main article in blue.
When finding the theoretical curve that best fits the data,
we fix the Raman factor to be the same as in the HH(0$^\circ$) configuration,
$Y^R_{VV}(45^\circ) = Y^R_{HH}(0^\circ) = 51450\ [\text{Hz/W}^2]^{1/2}$,
which should be the case since
$Y^R_{VV}(45^\circ) = C\mathcal{E}_0^2[A^E_{xyyx} +(A^E_{xxyy} +A^E_{xyxy}) ]/2 = Y^R_{HH}(0^\circ)$.
The best fitting curve, shown in Fig.~\ref{IM:Fig}~(b) of the article, is is obtained by choosing
$Y^E_{VV}(45^\circ) = MY^E_{VV}(0^\circ) (1.61 - i0.55)$,
which gives
$MY^E_{VV}(45^\circ) = [(147 \pm 10) + i(-50 \pm 4)]\ [\text{Hz/W}^2]^{1/2}$.

We could, however, have taken advantage of the tensorial relations and calculate via the expression
$Y^E_{VV}(45^\circ) = C\mathcal{E}_0^2[A^E_{xxxx} +A^E_{xyyx} +(A^E_{xxyy} +A^E_{xyxy}) ]/2$,
whose components we have already calculated,
yielding
$MY^E_{VV}(45^\circ) = [(188 \pm 15) + i(-12 \pm 4)]\ [\text{Hz/W}^2]^{1/2}$.
We attribute the discrepancy to the difficulty in extracting the correct $(A^E_{xxyy} +A^E_{xyxy})/A^E_{xxxx}$ ratio from the low HH(45$^\circ$) counts.
Since the counts are low, the uncertainty is most probably underestimated, and a lower
$(A^E_{xxyy} +A^E_{xyxy})/A^E_{xxxx}$ would better fit the VV(45$^\circ$) counts.
\end{document}